\documentclass[aps,pre,twocolumn,groupaddress,superscriptaddress,floatfix,
showpacs]{revtex4-1}

\usepackage[version=3]{mhchem}
\usepackage{subfigure}
\usepackage{graphicx}
\usepackage{comment}
\usepackage{wasysym}

\newcommand{\ra}{\mbox{$\rightarrow$}}
\newcommand{\dg}[1]{\mbox{$#1^\circ$}}

\graphicspath{{./eps_figures/}}

\begin{document}

\title{Logical and Arithmetic Circuits in Belousov Zhabotinsky
  Encapsulated Discs}

\author{Julian Holley}
\email[]{julian2.holley@uwe.ac.uk}
\homepage[]{http://uncomp.uwe.ac.uk/holley}
\affiliation{Faculty of Environment and Technology, University of the
  West of England, Bristol, England}


\author{Ishrat Jahan}
\affiliation{School of Life Sciences, University of the West of
  England, Bristol, England} 

\author{Ben {De~Lacy~Costello}}
\affiliation{School of Life Sciences, University of the West of
  England, Bristol, England} 

\author{Larry Bull}
\affiliation{Faculty of Environment and Technology, University of the
  West of England, Bristol, England}

\author{Andrew Adamatzky}
\affiliation{Faculty of Environment and Technology, University of the
  West of England, Bristol, England}

\date{\today}

\begin{abstract}

  Excitation waves on a sub-excitable Belousov Zhabotinsky (BZ)
  substrate can be manipulated by chemical variations in the substrate
  and by interactions with other waves.  Symbolic assignment and
  interpretation of wave dynamics can be used to perform logical and
  arithmetic computations. We present chemical analogs of elementary
  logic and arithmetic circuits created entirely from interconnected
  arrangements of individual BZ encapsulated cell like discs.
  Inter-disc wave migration is confined in carefully positioned
  connecting pores.  This connection limits wave expansion and unifies
  the input-output characteristic of the discs. Circuit designs
  derived from numeric simulations are optically encoded onto a
  homogeneous photo-sensitive BZ substrate.

\end{abstract}



\maketitle

\section{Introduction}


{\it Unconventional Computing}~\citep{UNCONVENTIONAL_2007} is a field
of study dedicated to exploring alternate computational strategies,
structures and mediums contrasting contemporary digital von Neumann
architecture machines~\citep{Neumann2005}.  Innate serial execution
and storage processing of these machines presents profound throughput
and energy restrictions, limiting practical application of some
classes of algorithms~\citep{Harel2003}.  Conversely computational
capable organic adaptive systems are innately parallel and
distributed.  Highly parallel and distributed computation can be
supported in chemical reaction-diffusion (RD)
systems~\citep{Adamatzky2005} such as the Belousov Zhabotinsky~(BZ)
reaction~\citep{Zhabotinsky1973}.  One bench mark of these systems is
the ability to replicate components used in conventional computation,
such as logic and arithmetic circuits. Logic gates are the physical
embodiment of Boolean logic operations that form the foundation for
digital computation.  Circuits of logic gates can be connected to
create machines capable of performing {\it `Universal
  Computation'}~\citep{Turing1936}.

The chemical implementation of logic gates were first described by
Rossler sustained in a bistable chemical medium~\citep{Rossler1974}.
Logic gates have also been demonstrated using chemical
kinetics~\citep{Hjelmfelt1991, Hjelmfelt1992, Hjelmfelt1993,
  Magnasco1997}.  The binary value assignment to the presence or
absence of individual RD waves modulated by the substrate geometry and
other waves presents an alternative approach.  Toth first described
logic gates with BZ waves along these lines~\citep{Toth1994,
  Toth1995}.  In that system RD waves travelled along thin capillary
tubes.  These basic principles have since led to numerous adaptations,
for example Steinbock implemented logic gates by `printing' a catalyst
of the BZ reaction onto a facilitating medium~\citep{Steinbock1995}.
In a simulated study Motoike presented (amongst other designs) logic
gates reliant upon geometric patterns of passive diffusion boundaries
embedded on an excitable field~\citep{Motoike1999}.  Sielewiesiuk also
created logical functions from using passive diffusion lines where
function arose at the intersection of perpendicular wave
channels~\citep{Sielewiesiuk2001}.  Adamatzky presented alternative
logic gate designs by combining the principles of collision based
computing on an precipitating chemical
substrate~\citep{Adamatzky2002b, Adamatzky2004b} and in-vitro
\citep{DeLacyCostello2005}.  Similar or derivations on the above
studies such as: \citep{Motoike2005} using the FitzHugh-Nagumo
model~\citep{FitzHugh1961, Nagumo1962} reaction, or \citep{Stone2008}
co-evolutionary system, or cardiac cell model on a
GPU~\citep{Scarle2009} and amongst others~\citep{Gorecki2009} have all
implemented logic functions or gates either in simulation, in-vitro or
both.

Current experiments with lipid coated vesicles of excitable media at
our (and our partners~\citep{neuneu2010}) laboratories have led to a
novel geometric BZ mediated approach to creating logic and arithmetic
functions.  Representative of a 2 dimensional approximation of BZ
vesicles, networks of interconnected discs containing a sub-excitable
BZ formulation can be arranged to create various circuits.  Logic \&
arithmetic circuits and polymorphic gates have been simulated in
homogeneous hexagonal and orthogonal networks~\citep{Adamatzky2011a,
  Adamatzky2011b} and non-homogeneous arrangements~\citep{Holley2011}.

In these simulations, geometric arrangements of discs are joined
together by small connecting pores.  The resulting pattern modulates
a homogeneous BZ Oregonator simulation grid that has adapted for
monochromatic photo-sensitivity~\citep{Kuhnert1986}.  The disc
boundaries are created from rings of high intensity light to emulate
an impenetrable exterior membrane.  Lower intensity light forms the
interior disc area and connecting pores adjusted to support travelling
waves.

Logically symbolic waves are able to traverse the network modulated by
interaction with pathways and other waves.  The disc interior can be
exploited for free space collision style
reactions~\cite{Adamatzky2002b} whereas the pore loci and efficiency
can compartmentalise the resulting reaction~\citep{Adamatzky2011a}.
Circuits have been created from logical sub assembles in orthogonal
and hexagonal networks~\citep{Holley2011, Adamatzky2011a}.  Function
density can be increased by circuits that include variations in
relative disc size, pore efficiency and connection
angles~\citep{Holley2011}.  Using this later technique we have
assembled a compact elementary arithmetic circuit.

Conceivably other cell geometries (such a grids of square units) could
compartmentalise the cell function similar to discs.  Nevertheless,
the curvature of the wave front in combination with the disc geometry
provides a natural and convenient unit and network structure.  In
comparison to our previous free space~\cite{Costello2005} or
channelled logic optically projected circuits~\cite{Costello2011} the
combination of the natural positively expanding wave curvature the
disc geometry promotes functional inter-disc stability and
connectivity.

Use of a photo-sensitive adapted version of the Oregonator model
permits a simple migration from simulation to experiment.  Circuit
designs from the simulation can be projected directly onto an actual
photo-sensitive BZ medium.  Furthermore, the same substrate is capable
of supporting different successive circuits.  Detected activity in
assigned output discs can trigger the projection of new alternate
circuits onto the same substrate area.  Those triggering waves can be
{\it captured} into the new circuit, effectively becoming inputs into
the new circuit.  Migration of our stabilising disc centric circuit
simulations onto a reusable chemical substrate is the primary subject
of this report.

\section{Methods}

Designs are initially created in simulation and then the resulting
geometry is then projected onto the BZ substrate.  This substrate is a
photo sensitive BZ system~\citep{Gaspar1983} with a \cf{Ru(bpy)3^3+}
catalyst immobilised on silica gel~\citep{Wang1999}.  The medium is
oscillatory in the dark with composition\ \footnote{Solution of catalyst
  free BZ medium~:-~\cf{NaBrO3$=0.36M$}, \cf{CH2(COOH)2$=0.0825M$},
  \cf{H2SO4$=0.18M$}, \cf{BrMA$=0.165M$} \& catalyst
  concentration~@~$0.004M$}. Global illumination levels are manually
adjusted to promote marginal excitation waves. Simultaneous multiple
wave initiations are optically instigated.  Removing all inhibiting
light promotes mass wave excitation across the BZ substrate.  Normal
light levels and experimental pattern are then restored with the
exception of small circular zones at desired initiation points.  Wave
fragments continue to emerge from these dark zones out into the
marginal zone of the disc body.

\subsection{Numerical simulation}

Numerical simulations are based on a 2 variable version of the
Oregonator model~\citep{Noyes1972} as a model of the BZ
reaction~\citep{Zaikin1970, Zhabotinsky1973} adapted for
photo-sensitive modulation of the \cf{Ru}-catalysed
reaction~\citep{Kuhnert1986}.

\begin{eqnarray}
  \frac{\partial u}{\partial t} & = & \frac{1}{\epsilon} (u - u^2 - (f
  v + \phi)\frac{u-q}{u+q})   + D_u \nabla^2 u \nonumber \\
  \frac{\partial v}{\partial t} & = & u - v \nonumber
\label{equation:oregonator}
\end{eqnarray}

Variables $u$ and $v$ are the local instantaneous dimensionless
concentrations of the bromous acid autocatalyst activator \cf{HBrO2}
and the oxidised form of the catalyst inhibitor \cf{Ru(bpy)3^3+}.
$\phi$ symbolises the rate of bromide production proportional to
applied light intensity.  Bromide \cf{Br^-} is an inhibitor of the
\cf{Ru}-catalysed reaction, therefore excitation can be modulated by
light intensity; high intensity light inhibits the reaction.
Dependant on the rate constant and reagent concentration $\epsilon$
represents the ratio of the time scales of the two variables $u$ and
$v$.  $q$ is a scaling factor dependent on the reaction rates alone.
The diffusion coefficients $D_u$ and $D_v$ of $u$ and $v$ were set to
unity and zero respectively. The coefficient $D_v$ is set to zero
because it is assumed that the diffusion of the catalyst is
limited. Inputs where instigated by perturbing the activator $u =
1.0$ centrally in small circular disc areas, $rad = 2$ simulation
points ($SP$).

\begin{table}[tbh!]
  \centering
  \begin{tabular}{ccl}
    Parameter & Value & Description\\
    \hline\noalign{\smallskip}
    $\epsilon$ & $0.022$ & Ratio of time scale for variables $u$ and $v$ \\
    $q$ & $0.0002$ & Propagation scaling factor \\
    $f$ & $1.4$ & Stoichiometric coefficient \\
    $\phi$ & $\star$ & Excitability level (proportional to light level)\\
    $u$ & $\sim$ & Activator \cf{HBrO2}\\
    $v$ &  $\sim$ & Inhibitor \cf{Ru(bpy)3^3+}\\
    $D_u$ & $1.0$ & Activator diffusion coefficient\\ 
    $D_v$ & $0$ & Inhibitor diffusion coefficient\\ 
    $\Delta x$ & $0.25$ & Spatial step\\
    $\Delta t$ & $0.001$ & Time step\\
    \hline
  \end{tabular}
  \caption{Kinetic and numerical values used in numerical simulations
    of (Eq.~\ref{equation:oregonator}). $\star \ \phi$ Varies between
    two levels, sub-excited ($L1$) and inhibited ($L2$),
    $\phi_{L1} = 0.076$, $\phi_{L2} = 0.209$}
  \label{table:oregonator_values}
\end{table}

Numerical simulations were achieved by integrating the equations
using the Euler-alternating direction implicit (ADI) method. 
method~\citep{Press1992} with a time step $\delta t = 0.001$ and a
spatial step $\delta = 0.25$.  Experimental parameters are given in
Tab.~\ref{table:oregonator_values}.

Networks of discs where created by mapping 2 different $\phi$ values
(proportional to light intensity) onto a rectangle of homogeneous
simulation substrate.  To improve simulation performance the rectangle
size was automatically adapted depending on the size of the network,
but the simulation point ($SP$) density remained constant throughout.
The excitation levels, $L1\ra L2$ relate to the partially active disc
interiors and non-active substrate.

Discs are always separated by a single simulation point ($1~SP$) wide
boundary layer.  Connection pores between discs are created by
superimposing another small {\it link} disc at the point of connection
(typically a $2\ra 6~SP$ radius), simulation points have a
$\mbox{1:1}$ mapping with on screen pixels.  The reagent
concentrations are represented by a red and blue colour mapping; the
activator, $u$ is proportional to red level and inhibitor, $v$
proportional to blue.  The colour graduation is automatically
calibrated to minimum and maximum levels of concentration over the
simulation matrix.  The background illumination is mono-chromatically
calibrated in the same fashion proportional to $\phi$, white areas are
inhibitory and dark areas excited (for monochromatic reproductions,
blue ($v$) has been suppressed and red ($u$) appears as light gray,
colour versions online).

Wave fragment flow is represented by a series of superimposed time
lapse images (unless stated otherwise), the time lapse is 50
simulation steps. To improve clarity, only the activator ($u$) wave
front progression is recorded.  Figure~\ref{figure:time_lapse_example}
illustrates the same wave fragment in both colour map ($u$ \& $v$) and
final composite time lapse version ($u$).

\newcommand{\tw}{\textwidth/2}

\begin{figure}[!ht]
\centering
\subfigure[]{\includegraphics[width=0.15\tw]{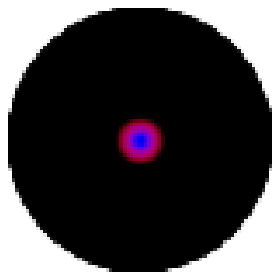}}
\subfigure[]{\includegraphics[width=0.15\tw]{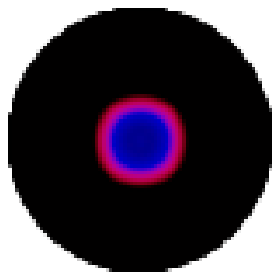}}
\subfigure[]{\includegraphics[width=0.15\tw]{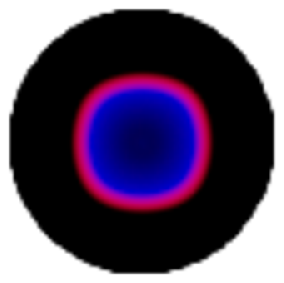}}
\subfigure[]{\includegraphics[width=0.15\tw]{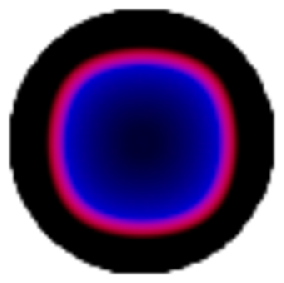}}
\subfigure[]{\includegraphics[width=0.15\tw]{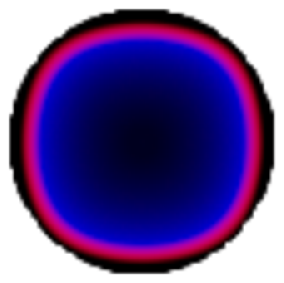}}
\subfigure[]{\includegraphics[width=0.15\tw]{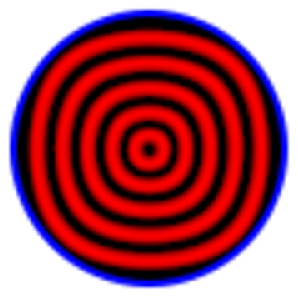}}
\caption{(Colour online) Example the time lapse image creation.  The
  image shown in (f) is the accumulation of successive images shown
  from central wave initiation in (a) to extinction in (e).  Time
  lapses periods are 50 time steps and the refractory tail of the
  inhibitor ($u$) shown in blue (gray inner shadow of propagating ring
  [(a)\ra(e)]) is not shown in the time lapse image (f) to improve
  clarity.}
\label{figure:time_lapse_example}
\end{figure}

Inputs are created by perturbing a small circular area of the
activator ($u$) set to a value of $1.0$ with a radius of $2~SP$ in the
center of the disc.  In addition to explicit labelling, all discs
representing inputs and outputs are also highlighted with a blue and
green border respectively.

\begin{figure}[!ht]
\centering
\subfigure[Relative disc sizes]{\includegraphics[width=0.3\tw]{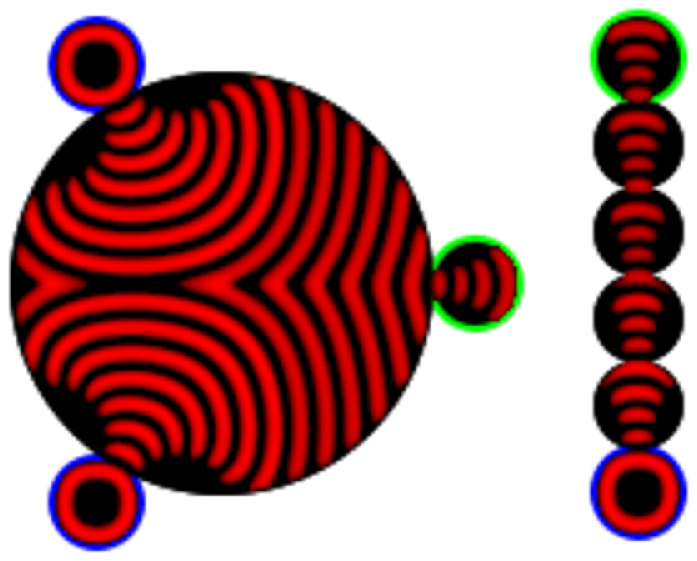}}
\subfigure[Connection angle]{\includegraphics[width=0.3\tw]{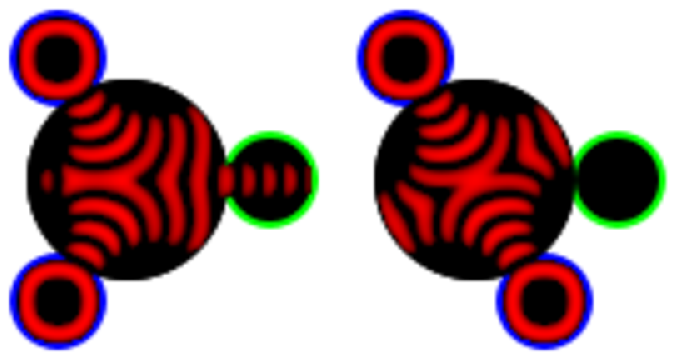}}
\subfigure[Connection efficiency]{\includegraphics[width=0.3\tw]{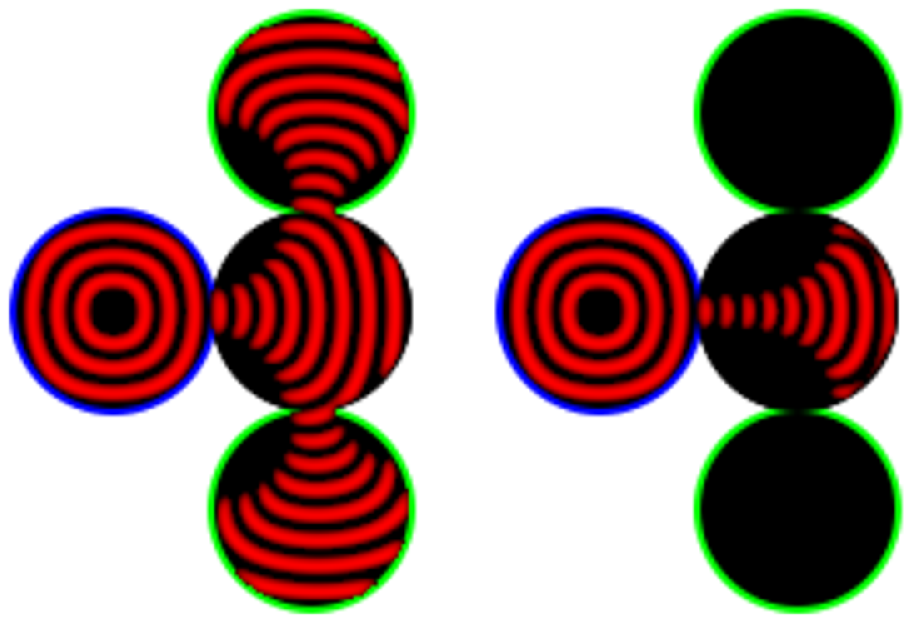}}
\caption{(Colour online) (a) Discs as reaction vesicles in the LHS
  network or communication channels in the RHS network. (b) The effect
  of connection angle, the two input signals combine in the left
  network to produce an output.  Adjusting the angle of the lower
  input in the right network alters the result of the collision and no
  output is produced.  (c) The effect of connection efficiency.  Large
  aperture ($6~SP$) connection in the left network results in a broad
  spreading beam. Conversely a smaller ($4~SP$) aperture connection in
  the right network creates a narrow beam wave.}
\label{figure:signal_modulation}
\end{figure}

In a previous study we have shown that logic circuits can be created
with uniform discs arranged in hexagonal
networks~\citep{Adamatzky2011c}, hexagonal packing being the most
efficient method of sphere (disc) packing.  Further opportunities to
modulate wave fragment behaviour are presented when disc size,
connection angle and connection efficacy are combined in
non-homogeneous networks.  Disc size can be adjusted to permit or
restrict internal wave interactions, producing either larger reaction
vessel discs or smaller communications
discs~(Fig.~\ref{figure:signal_modulation}a).  Connection angle
between discs can be used to direct wave
collisions~(Fig.~\ref{figure:signal_modulation}b) and connection
efficiency can effect the wave
focus~(Fig.~\ref{figure:signal_modulation}c).

\subsection{Experimental}

\begin{figure}[!ht]
\fbox{\includegraphics[width=0.4\textwidth]{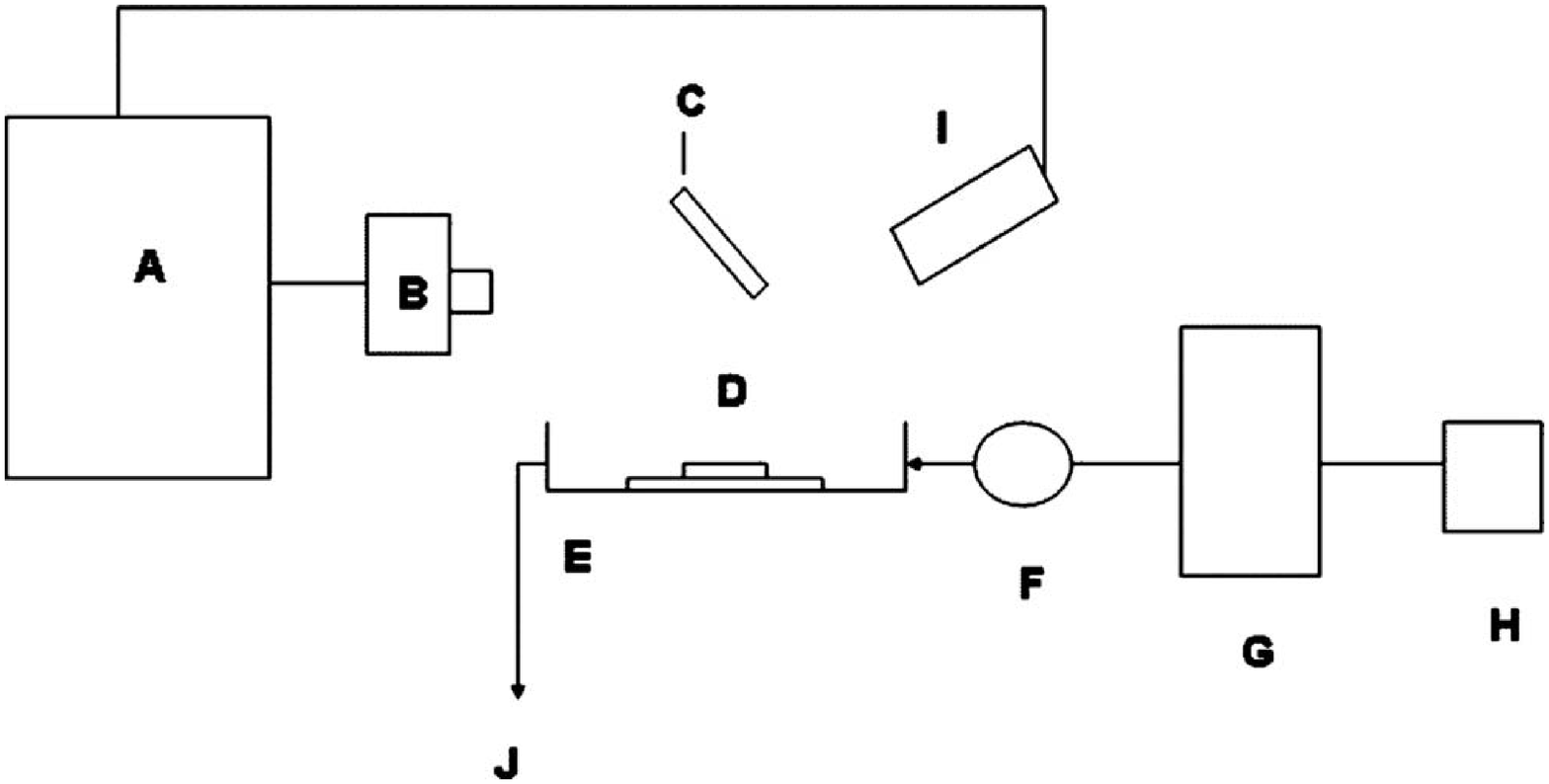}}
\caption{Experimental setup: A light sensitive catalyst
  \cf{Ru(bpy)3^3+} loaded silica gel is immersed in catalyst free BZ
  reaction solution in a thermostatted (G) Petri dish (E). Peristaltic
  pump (F). Reactor with thermostatted reaction solution and to remove
  effluent (J). The reaction solution reservoir (H) is kept in an ice
  bath. The heterogeneous network on the surface of the gel (D) is
  constructed by the projection (B) of a disc pattern generated by a
  computer (A) via a mirror assembly (C). Images captured with a
  Digital camera fitted with a $455nm$ narrow bandpass interference
  filter (I).}
\label{figure:experimental_setup}
\end{figure}

A Sanyo proxtrax multiverse projector was used to project the design
outline of several simulated circuit designs from~\cite{Holley2011}.
Wave activity was captured using a Lumenera infinity 2 USB 2.0
scientific digital camera. The open reactor was surrounded by a water
jacket thermostatted at \dg{20}C. Peristaltic pumps were used to pump
the reaction solution into the reactor and remove the effluent
(Fig.~\ref{figure:experimental_setup}).

Sodium bromate, sodium bromide, malonic acid, sulphuric acid,
tris(bipyridyl) ruthenium(II) chloride, $27\%$ sodium silicate
solution stabilised in $4.9M$ sodium hydroxide\footnote{Purchased from
  Aldrich (U.K.)} and used as received unless stated otherwise.
\cf{Ru(bpy)3^3+} was recrystalised from the choloride salt with
sulphuric acid.  To create the catalyst loaded gels a thin layer
chromatography pre-coated plates silica gel with $254nm$ fluorescent
indicator on glass was cut into $5cm \times 5cm$ and placed in a
solution of $0.9ml$ of $0.025M$ \cf{Ru(bpy)3^3+} and $12ml$ of deionised
water in a Petri dish for 12 hours. Gels were washed in deionised
water to remove by products before use.

The catalyst free reaction mixture was freshly prepared in a $30ml$
continuously-fed stirred tank reactor (CSTR), which involved the in
situ synthesis of stoichiometric bromomalonic acid from malonic acid
and bromine generated from the partial reduction of sodium
bromate. This CSTR in turn continuously fed a thermostatted open
reactor with fresh catalyst-free BZ solution in order to maintain a
nonequlibrium state. The final composition of the catalyst-free
reaction solution in the reactor was: $0.42M$ sodium bromate, $0.19M$
malonic acid, $0.64M$ sulphuric acid and $0.11M$ bromide. The residence
time was $30$ minutes.  

The spatially distributed excitable field on the surface of the gel
was achieved by the projection of the disc pattern from the
simulations. The light intensity of discs was controlled by computer.
The pattern was projected onto the catalyst loaded gel via a data
projector.  Every $10$ seconds, a light level of $5.7mW/cm^2$ for
$10ms$ during which time an image of the BZ fragments on the gel was
captured. The purpose of this was to allow activity on the gel to be
more visible to the camera.  Captured images were processed to
identify chemical activity. This was done by differencing successive
images to create a black and white image. The images were cropped and
layered to show progression of a single image finally the disc
boundary was superimposed on the images to aid analysis of the
results.

\section{Results}

We report on a selection of experiments that successfully migrated
from simulation; a diode, {\sc nand} gate, {\sc xor} gate and an
simple arithmetic adder circuit~\citep{Holley2011}.  In the following
images the inverse monochromatic wave progression (left column) is
illustrated adjacent to the respective colour simulation analogue
(right column).  Although the parameterisation of the simulation
creates different wave dynamics the projected geometry in each case is
proportionally identical resulting in the same functionality.

\subsection{Diode}

\begin{figure}[!ht]
\includegraphics[width=0.30\textwidth]{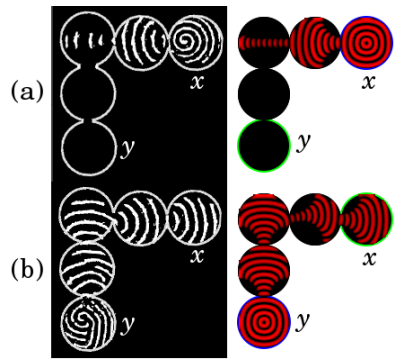}
\caption{(Colour online) Diode junction ($x\ra y = 0$, $y\ra x = 1$).
  Left column sequences show wave (signal) development compared
  alongside analogous numerical simulations on the right.  (a) Signal
  propagates horizontally ($x\ra y$).  A narrow band pore ($4~SP$) at
  the $2^{nd}$ (right angle) junction prohibits propagation downwards
  to $y$. (b) Signal propagates vertically ($y\ra x$).  A broad band
  pore ($6~SP$) at the $2^{nd}$ (right angle) junction connection
  permits the signal to expand horizontally towards $x$.  Simulation
  to gel projection ratio is \mbox{$56~SP:10mm~\diameter$}
  (\mbox{Mov.~1a\&b~\citep{p2sup2011}}).}
\label{figure:diode_juction}
\end{figure}

The diode junction constructed with BZ discs operates by
simultaneously exploiting a right angle junction and asymmetric pore
size (Fig.~\ref{figure:diode_juction}).  A wave fragment ($y$) travels
horizontally from \mbox{right\ra left}
(Fig.~\ref{figure:diode_juction}a). Wave fragments cannot survive when
the fragment size drops below some critical level~\citep{Kusumi1997}.
A constricted pore of $(4~SP)$ at the perpendicular junction reduces
the wave size close to termination in the proceeding disc.  Wave
development is momentarily marginal before slowly recovering.  Delay
in wave development prohibits the fragment entering the vertical
connection (Movies \mbox{Mov.~1a\&b~\citep{p2sup2011}}).  Conversely
the opposing vertical fragment ($y$) enters the junction through a
broad pore $(6~SP)$.  The resultant increased junction wave seed
permits the fragment to develop more rapidly and dissipate into the
horizontal discs line (Fig.~\ref{figure:diode_juction}b). Therefore
signal propagation only progresses from vertical to horizontal ($x\ra
y$).  Discs in the simulation have a diameter of $56~SP$ which results
in a projected gel disc of $10mm~\diameter$.  This diode junction has
been used to inject a unidirectional wave fragments into a type of
simulated 1~bit memory where unidirectional waves indefinitely
circulate around a loop of discs until reagent exhaustion or collision
with an opposing wave~\citep{Holley2011}.

\subsection{{\sc nand} logic gate}

\begin{figure}[!ht]
\includegraphics[width=0.38\textwidth]{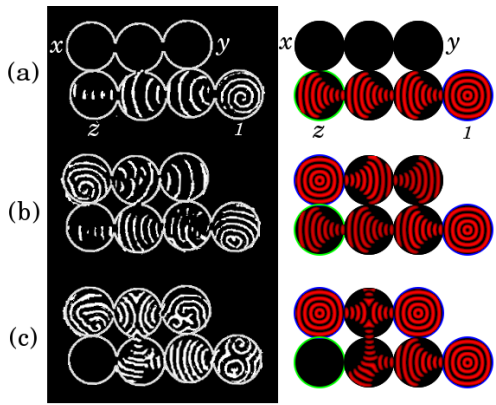}
\caption{(Colour online) Two input {\sc nand} gate ($z =
  \overline{x\bullet y}$).  Left column sequences show wave (signal)
  development compared alongside numerical simulations on the
  right. (a) $z = 1, (x, y) = [0, 0]$. (b) $z = 0, (x, y)=[0, 1] | [1,
  0]$. (c) $z = 0, (x, y) = [1, 1]$. Pores of $6~SP$ interconnect
  $56~SP \diameter$ discs with simulation to gel projection ratio of
  \mbox{$56~SP:10mm~\diameter$} (\mbox{Mov.~2a\ra
    c}~\citep{p2sup2011}).}
\label{figure:nand_gate}
\end{figure}

A {\sc nand} gate design is shown in Fig.~\ref{figure:nand_gate} and
animated in movies \mbox{Mov.~2a\ra c}~\citep{p2sup2011}.  The gate is
a conjunction of an {\sc and} gate and an {\it inverter} or {\sc not}
gate.  The {\sc not} gates operates horizontally in the bottom four
discs.  A logical {\sc truth} or `1' (supply wave) is always presented
simultaneously with the inputs $x\ \&\ y$.  In the case of $(x,y) =
(0,0)$ a wave propagates horizontally unimpeded terminating in the
output disc $z$, $(z = 1)$ (Fig.~\ref{figure:nand_gate}a).
Presentation of input cases $(x,y) = (1,0)\ |\ (0,1)$ also have no
effect on the progression of the supply wave and in both cases $z = 1$
(Fig.~\ref{figure:nand_gate}b).  In the final instance $(x,y) = (1,1)$
the inputs collide in the central disc resulting in a perpendicular
ejection wave.  The downward ejection wave fragment propagates into
the horizontal {\sc not} line below and deflects and causes the
extinction of the supply wave leading to a negated output $(z = 0)$
(Fig.~\ref{figure:nand_gate}c).  The {\sc nand} gate is significant
because of its universal applicability in the construction of all
other logic gates.  This design is illustrative of the modular
facilitation of BZ disc networks creating a {\sc nand} gate from the
conjunction of an {\sc and} and a {\sc not} gate.

\subsection{{\sc xor} logic gate}

\begin{figure}[!ht]
\includegraphics[width=0.38\textwidth]{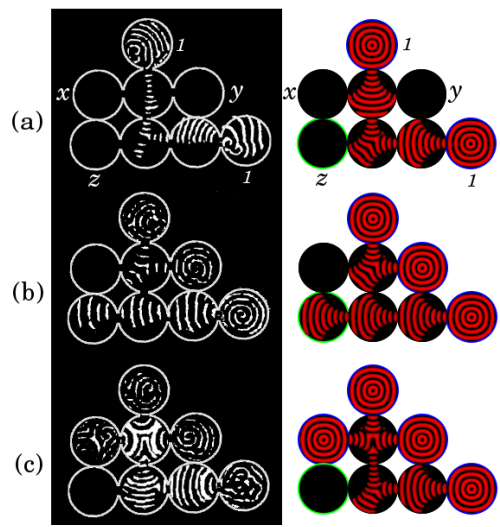}
\caption{(Colour online) Two input {\sc xor} gate ($z = x\oplus
  y$). Left column sequences show wave (signal) development compared
  alongside numerical simulations on the right for the input sets (a)
  $z = 0, (x, y) = [0, 0]$. (b) $z = 1, (x, y) = [0, 1] | [1, 0]$. (c)
  $z = 0, (x, y) = [1, 1]$.  Pores of $6~SP$ interconnect $56~SP
  \diameter$ discs with simulation to gel projection ratio of
  \mbox{$56~SP:10mm~\diameter$} (\mbox{Mov.~3a\ra
    c}~\citep{p2sup2011}).}
\label{figure:xor_gate}
\end{figure}

Figure.~\ref{figure:xor_gate} demonstrates the {\sc xor} function. As
with the proceeding {\sc nand} gate, a logical {\sc truth} (`1') is
supplied in two discs (centre top and bottom right) in synchronisation
with the two inputs ($x\ \& \ y$).  Presentation of $(x,y) = (0,0)$
allows the vertical supply wave to deflect the horizontal supply wave
resulting in logical false output $z = 0$
(Fig.~\ref{figure:xor_gate}.a).  Complimentary inputs $(x,y) = (0,1)\
\&\ (1,0)$ deflect the vertically travelling supply wave allowing the
horizontal supply wave reaching the output disc $(z = 1)$
(Fig.~\ref{figure:xor_gate}.b).  Finally the input presentation of
$(x,y) = (1,1)$ cancels the effect of the asymmetric collision of the
vertical supply wave again deflecting the horizontal wave path $(z =
0)$ (Fig.~\ref{figure:xor_gate}.c) (\mbox{Mov.~3a\ra
  c}~\citep{p2sup2011}).

Using this design strategy we have constructed other simulated gates
({\sc not}, {\sc and}, {\sc or}, {\sc nxor}) and with combinations the
we have also simulated simple memory and arithmetic
circuits~\citep{Holley2011}.

\subsection{Composite one bit half adder circuit}

One of the building blocks of electronic digital arithmetic circuits
is the 1 bit half adder ({\sc ha}).  Two {\sc ha} circuits can be
connected to make a full 1 bit adder.  One bit adders can then be
repeated connected to make an $n$ bit adder.  A successful
construction of a {\sc ha} is therefore demonstrative of arithmetic
capabilities of this scheme.  The {\sc ha} circuit takes 2 inputs; the
$n$ bit $(x)$ in conjunction with a carry $(y)$ and produces two
outputs, the sum $(S)$ and carry $(C)$.  A {\sc ha} can be
constructed from a combination of two logic gates the {\sc xor} and
{\sc and} gate.  There are two inputs ($x$ \& $y$) and two outputs
($S$ \& $C$), the binary sum ($S$) of $x$ \& $y$ is achieved by the
{\sc xor} gate ($S = x \oplus y$) and inability of the configuration
(overflow) to present the $1 + 1$ input is achieved with a carry ($C$)
output, ($C = x\bullet y$).

We have conceived several simulated half adders and full adders
circuits arranged in orthogonal~\citep{Holley2011} and
hexagonal~\citep{Adamatzky2011a} networks where {\sc and} \& {\sc xor}
gates form building blocks in a larger circuit.  Nevertheless it is
possible to increase functional density by adding flexibility.  With
the exception of the previous diode, the above logic gates are
constructed with discs connected with a uniform orthogonal network and
connection pores.  More efficient designs are possible by permitting
flexibility in terms of disc size, connection pore (see diode) and
inter disc connection angle.

\begin{figure}[!ht]
\includegraphics[width=0.38\textwidth]{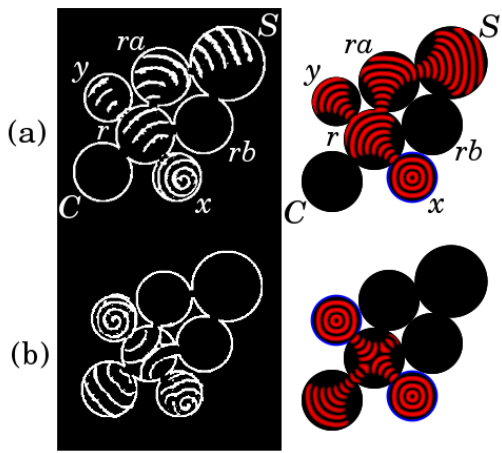}
\caption{(Colour online) One bit half adder circuit ($S = x\oplus y,\
  C = x\bullet y$).  The central reactor disc perform both {\sc xor}
  and {\sc and} logic functions to create the desired logic. (a) {\sc
    xor} component, $S = 1, C = 0, (x,y) = [0,1]|[1,0]$. (b) {\sc and}
  component, $S = 0, C = 1, (x,y) = [1,1]$.  Pores of $4~SP$
  interconnect discs $56~SP \diameter$ disc, scaled by $1.2$ for $S$
  and $0.8$ for $x \& y$ with simulation to gel projection ratio of
  $\mbox{56~SP:10mm~\diameter}$. The circuit exploits 3 types of wave
  modulation, connection angle, disc size and pore efficacy)
  (\mbox{Mov.~4a\&b~\citep{p2sup2011}}).}
\label{figure:ha_gate}
\end{figure}

Figure~\ref{figure:ha_gate} illustrates a compact {\sc ha} circuit design
where the {\sc xor} \& {\sc and} gate are combined in a central single
{\it reactor} disc $(R)$.  Interconnecting discs $(r1, r2)$ between
the central reactor disc guide wave fragments to the $S$ output to
create the {\sc xor} feature $(xy(1,0)\ \&\ xy(0,1)$\ra $S = 1,\ C =
0)$.  The {\sc and} feature is created by the perpendicular wave
ejection fragments directly into the $(C)$ $(xy(1,1)$\ra $S = 0,\ C =
1)$ (\mbox{Mov.~4a\&b~\citep{p2sup2011}}).

\section{Discussion}

In much the same way that electronic circuits are designed and
constructed we have created an alternative {\sc ha} circuit to that
presented above.  In that instance the {\sc ha} circuit is formed from
the conjunction of an {\sc and} gate unit and a {\sc xor} gate unit.
Recursive connections too have led to the creation of memory and
sequential circuits that can be combined with diodes and logic gates
to create other circuits. Beyond the limitations of our own laboratory
resources we foresee no reason preventing successful migration of
these additional circuits.

Although these circuits share some similarity to their electronic
counterparts, there are differences.  Reagents, the chemical analog
of the {\it power supply} is embedded and distributed throughout the
circuit substrate and not supplied from one source.  A circuit can
thus activate for short periods in absence of any reagent refreshment.
The vector component of the reaction wave presents another interesting
contrast.  Temporally separated waves can share a common channel
without interference.  It is therefore possible to create a
`cross-roads' junction where wave signals travelling in different
directions can share the same substrate without
interference~\citep{Holley2011}.  Furthermore the optical circuit
projection onto a homogeneous substrate permits substrate reuse with
preservation current signals.  One operating circuit projection can be
instantaneously replace with another circuit.  Waves in target
locations could trigger circuit changes and conceivably also
incorporate other existing wave signals into the {\it new} circuit.

At a scale where excitation waves are visible with the human eye this
(BZ) reaction diffusion process is relatively slow.  In these
experiments discs are $\sim 10mm$ in diameter and waves propagate at
approximately $\sim 0.5mms^{-1}$ resulting in a $20 s$ propagation
delay per disc.  Wave front geometry is an emergent property of the
chemical composition and substrate, that develops independently to the
initiation geometry.  Reducing the disc physical dimensions therefore
requires an proportionate increase in chemical kinetics to maintain
functional equilibrium.  For example; if the chemical kinetics could
be adapted to reduce the wave front size, then a reduction in diameter
to that of a typical neuron from ($10mm\ra 0.004mm$) could at least
produce a linear adjustment in disc propagation delay $20 / 2500 = 8
ms$; in comparison the refractory phase of a typical neuron is $\sim
1ms$.

This research is an exploratory component within a wider collaborative
project that aims to create functional networks of lipid encapsulated
BZ vesicles~\citep{neuneu2010}.  The lipid membrane and the non-linear
oscillatory nature of the BZ medium encodes some of the features
apparent in biological information processing; there are systematic
analogies between electro-biochemical neurons and BZ discs, such as
pores\ra~synapses and chemical\ra~electrical excitation and
refraction.

\section{Future work}



To demonstrate the computational abilities of BZ encapsulated discs we
have combined Boolean information representation and geometric
collision style manipulation.  The essential components required for a
universal computer could be created with this combination.
Nevertheless other modes of information representation and
manipulation are possible in these networks.  In an oscillatory mode,
information could be coded in pulse phase relationships and
manipulated by interconnected temporal associations, reminiscent of
natural neural information processing.  Information is processed
probabilistically along the edge of the BZ instability threshold
through interconnected membrane pores.  Larger reactor chambers could
possibly be connected by strings of smaller discs or vesicles.

Our explorations have so far been restricted in 2 dimensions.  Discs
have been substituted for cross sections of interconnected vesicles.
In future studies we plan to extend our simulations into the third
dimension.  We speculate that similar modulation of wave {\it cones}
could also be possible leading to 3D vesicle computational or adaptive
circuits.

Designing anything beyond all but the simplest devices in such schemes
as is the case with neural networks, can be uncertain and complex.  We
intend to create discs capable of adaptation in order to learn
solutions as well as employing evolutionary strategies to search for
static structural solutions.  Currently we are exploring an
evolutionary strategy to replicate the functionality of our manual
designs of logic gates and arithmetic circuits.  Success there opens
the possibility on solving computational tasks for which solutions are
currently protracted in conventional systems.

\section{Conclusion}

Concomitant within our long term project aims we have developed a
prototype information processing system from interconnected
arrangements of BZ encapsulated discs as analogs of BZ vesicles.
Expanding excitation waves sustained in a sub-excitable BZ substrate
represent discrete quanta of information.  The inter-connecting disc
pores have a stabilising effect on the waves as they propagate from
one disc to another.  Connections, pore efficiency, connection angle
and relative disc size are used to manipulate waves to create explicit
chemical information processing devices.

Networks of these units share some features apparent in biological
information processing.  Whilst inter-neuron communication is
predominately electrical, modulation of that activity is chemical. In
the case of individual neurons modulation dominates at the synaptic
junction.  The membrane pore between two vesicles can be considered a
simple analog of the synapse, a small contact area that can modulate
signals in between vesicles.  Similarly the electric upstate firing
and downstate quiescence of neural signalling is an analogue of
chemical excitation and refraction.

\section{Acknowledgements}

This work is funded under $7^{th}$ FWP FET European project 248992.
We thank the project coordinator Peter Dittrich and project partners
Jerzy Gorecki \& Klaus-Peter Zauner for their inspirations and useful
discussions~\citep{neuneu2010}. The authors also acknowledge the EPSRC
grant no. \mbox{EP/E016839/1} for support of Ishrat Jahan.


%


\end{document}